\documentclass[aps,pra,prabib,twocolumn,showpacs,nofootinbib]{revtex4}
\usepackage{graphicx} \usepackage{amsmath} \usepackage{amssymb}
\usepackage{amsfonts} \usepackage{bm}

\begin{document}

\newcommand{\be}{\begin{equation}} \newcommand{\ee}{\end{equation}}
\newcommand{\bea}{\begin{eqnarray}}\newcommand{\eea}{\end{eqnarray}}

\title{Physical realization for Riemann zeros from black hole physics}

\author{Pulak Ranjan Giri} \email{pulakranjan.giri@saha.ac.in}

\affiliation{Theory Division, Saha Institute of Nuclear Physics,
1/AF Bidhannagar, Calcutta 700064, India}
\affiliation{Department of Physics and Astronomy,
McMaster University, Hamilton, Ontario, Canada L8S 4M1}

\begin{abstract}
According to a conjecture attributed to Polya and Hilbert,
there is a self-adjoint operator whose eigenvalues are the the
nontrivial zeros of the Riemann zeta function.
We show that the near-horizon dynamics of a massive scalar field in
the Schwarzscild black hole spacetime, under a reasonable boundary
condition, gives rise to normal mode frequencies  that coincide with the
nontrivial Riemann zeros. In achieving this result, we exploit the
Bekenstein  conjecture of black hole area quantization, and
argue that it is responsible for the breaking of the continuous scale
symmetry of the near horizon dynamics into a discrete one.
\end{abstract}

\pacs{03.65.Ge, 02.10.De, 04.70.-s, 04.70.Dy}

\date{\today}

\maketitle

{\it I. Introduction.---}
The Riemann zeta function is of great importance in number theory, and
is of interest to physicists because of its connection to quantum
chaos and random matrix theory~\cite{bogo, berry, snaith}.
Starting with the Dirichlet series
\begin{equation}
\zeta(s)=\sum_{n=1}^{\infty}\frac{1}{n^s}~,
\label{dir}
\end{equation}
which converges for ${\cal R}e~s>1$, Riemann~\cite{riemann} analytically continued it
over the whole complex plane by defining an appropriate contour
integral that reduced to Eq.(\ref{dir}) for ${\cal R}e~s >1$.
Writing $s=(s_1+is_2)$, the only pole is on the real axis at $s_1=1$,
and the ``trivial zeros'' are on the real axis at negative
even integer values of $s_1$. According to Riemann hypothesis, the {\it
  only} other zeros of $\zeta(s)$, infinite in number, lie in
complex conjugate pairs on the line
$s_1=1/2$ in the complex $s$-plane. This has been numerically tested over extensive
samples by Odlyzko~\cite{od}, but has not been proven mathematically \cite{hardy}.
According to a conjecture attributed to Polya and
Hilbert, these zeros on the upper half of the complex
plane may be the real eigenvalues of a self-adjoint operator in quantum
mechanics. Previous work on this topic are due to, amongst others,
Hua Wu and Sprung~\cite{sprung}, Berry and Keating~\cite{berry}, Connes~
\cite{connes}, and Sierra and Townsend~\cite{sierra}.

The purpose of this paper is to suggest such an example from
black-hole (BH) physics. Specifically, we consider the Klein-Gordon
spectrum of a massive scalar field just outside the horizon of a
Schwarzschild BH~\cite{hooft}. It is well known that the near-horizon dynamics of
the scalar field is described by a scale-invariant inverse-square potential with a
coupling strength that is a function of the eigenvalues of the scalar
field~\cite{gupta}. This is quite different from a nonrelativistic self-adjoint
operator with a constant coupling. The scale invariance is broken by
other terms in the
Hamiltonian at larger distances from the horizon. The correspondance
of the eigenvalues of the scalar field with the Riemann zeros depends
crucially on the assumption of scale invariance with energy-dependent
coupling, and therefore refers to only a subset of the deeply bound
eigenvalues that come from the near-horizon dynamics. There are two
other important ingredients in obtaining the desired result. The
inverse square potential is cut off by a brick-wall as introduced
by 't Hooft~\cite{hooft} very near the horizon, but we demand that the final
spectrum should not depend explicitly on the brick-wall
radius. Lastly, we make the crucial assumption that the area of the BH
is quantized in units of the Planck length squared. In order to preserve
 this quantisation under a scale transformation in D-spatial dimensions
$(D\geq 1)$, the continuous scale symmetry breaks into a discrete one.
In section II we analyze the massive scalar field equation in
Schwarzschild black hole spacetime. In section III the near-horizon
approximation is made that gives a scale invariant
Hamiltonian. The normal modes which have been found are identified
with the nontrivial Riemann zeros  of the zeta function. Finally we
conclude in section IV with a discussion.

{\it II. Scalar field in Schwarzschild black hole spaceime.---}
As mentioned in the introduction, we go beyond the
non-relativistic self-adjoint operator to obtain our desired result.
We begin our discussion with  Schwarzschild black hole, which is a
$(D+1)$-dimensional spacetime obtained from the vacuum Einstein Equation.
For simplicity we consider a $(3+1)$-dimensional  spacetime  defined by the metric
\begin{eqnarray}
ds^2= - Ndt^2+ N^{-1}dr^2 + r^2d\Omega^2\,. \label{metric}
\end{eqnarray}
where
\begin{eqnarray}
N= 1-\frac{2M}{r}\,,
\end{eqnarray}
$M$ is  the mass of the black hole and $d\Omega^2$ is the metric over $S^2$.
The dynamics of a  massive scalar field of mass $m_S$ in this black
hole spacetime is given by the Klein Gordon equation
\begin{eqnarray}
\frac{1}{\sqrt{-g}}\partial_\mu(\sqrt{-g}g^{\mu\nu}\partial_\nu\Psi)=
m_S^2\Psi\,.
\label{Klein}
\end{eqnarray}
Unlike the BTZ black hole~\cite{banados, giri2}, this field equation is hard
to solve exactly. However in order to get an idea
of the normal modes for the scalar field, a careful look is needed for the
corresponding near-horizon approximation of the eigenvalue equation.
Because of spherical symmetry, we can write the $l$-th partial wave-function for the normal  mode $\omega$ as
\begin{equation}
\Psi(r,\theta,\phi;t)= \frac{1}{\sqrt{r-r_+}}
\mathcal{F}_\omega^{(l)}(r) e^{-i\omega_l
  t}Y_{lm}(\theta,\phi)~.
\end{equation}
In the above, $r_+=2M$ is the radius of the horizon, and $\omega_l$
the eigenenergies of the scalar field. Substituting
the ansatz in  (\ref{Klein})  the radial eigenvalue equation is obtained
to be
\begin{eqnarray}
H_{BH}\frac{1}{\sqrt{r- r_+}}\mathcal{F}_{\omega}^{(l)}(r)=0\,. \label{radial}
\end{eqnarray}
The operator  $H_{BH}$ in explicit form is written as
\begin{eqnarray}
\nonumber H_{BH} &=&r^2N\frac{d}{dr}\left(r^2N\frac{d}{dr}\right) \\
&+& \left(\omega_l^2r^4- m_S^2r^4N-l(l+1)r^2N\right) \,.
\label{loperator}
\end{eqnarray}
Henceforth we consider only the $l=0$ eigenstates that are
non-degenerate, and are relevant for our purpose. For simplicity, the
$l=0$ quantum number is supressed from now on.
Since (\ref{radial}) cannot be solved exactly, we proceed to obtain
some information from the near-horizon approximation.
In the spirit of 't Hooft~\cite{hooft}, to avoid divergences in free energy
at the horizon, we apply a ``brick wall'' boundary condition that
forces the scalar field to go to zero at the radius $r=r_+ + r_H$ where $r_+ +  r_H$
is slightly larger than the horizon radius $r_+=2M$. Our final result
is independent of $r_H$.

{\it III. Near-horizon dynamics.---}
In this section we consider near-horizon dynamics of a scalar field in
Schwarzschild  black hole spacetime. In terms of the near-horizon
coordinate $z=(r- r_+)$, the radial equation in the near-horizon limit  is given by
\begin{eqnarray}
\tilde{H}_{BH}\tilde{\mathcal{F}}_{\omega}(z)= 0\,,\label{latz01}
\end{eqnarray}
where
\begin{eqnarray}
\tilde{H}_{BH}= \frac{d^2}{dz^2}+ \left(\frac{1}{4}+ 4M^2\omega^2\right)\frac{1}{z^2} \,.\label{latz0}
\end{eqnarray}
Terms of  $\mathcal{O}(1/z)$ or lower have been neglected in this limit.
In general the operator like (\ref{latz0}) occurs in many branches of non-relativistic physics also, but the crucial difference is the presence of the scalar field frequency $\omega$ as the coefficient of inverse square term. This will in fact help us achieve our goal.   In order to show the relevance of (\ref{latz0})
to the Riemann zeros, let us now calculate  quantum mechanical  spectrum of
(\ref{latz01}).
There are  two independent  short-distance solutions of the form
\begin{eqnarray}
\nonumber \tilde{\mathcal{F}}^{I}_{\omega}(z) & =&
z^{1/2-iE}\,,\label{l1}\\
\tilde{\mathcal{F}}^{II}_{\omega}(z) &=&
z^{1/2+iE}\,, \label{latz0s}
\end{eqnarray}
where
\begin{equation}
\tilde{\mathcal{F}}_{\omega}(z)\sim Lim_{z\rightarrow 0}
{\mathcal{F}}_{\omega}(z),
\end{equation}
and
\begin{equation}
E=2M\omega~.
\label{short}
\end{equation}
The   most general short-distance  solution
is a linear sum of both,
\begin{eqnarray}
\tilde{\mathcal{F}}_{\omega}(z)= A_1 \tilde{\mathcal
  {F}}^{I}_{\omega}(z)
+ A_2 \tilde{\mathcal{F}}^{II}_{\omega}(z)\,,
\label{generaal1}
\end{eqnarray}
where $A_1$ and $A_2$ are constants.
Note, however, that $\tilde{H}_{BH}$ given by Eq.(\ref{latz0}) is a
scale invariant operator  and in general it  does not have any bound
state solutions in the interval $z\in[0,\infty]$ because the solutions
are not normalizable on the half line. However the asymptotic behavior
of the radial equation  (\ref{radial}) will make $\mathcal{F}_\omega(z)$ normalizable at
large distance. Restricting the dynamics  in a finite interval may
therefore give rise to bound state solutions.
In the interval $z\in[r_H,L]$ the operator (\ref{latz0}) will be
symmetric in a domain $D(\tilde H_{BH})$ if it satisfies  the condition
\begin{eqnarray}
\nonumber
&& \int_{r_H}^{L}\psi(z)^* \tilde H_{BH}\phi(z)dz - \int_{r_H}^{L}\left(\tilde H_{BH}\psi(z)\right)^* \phi(z)dz\\
\nonumber &=&\psi(L)^*\phi'(L)- \psi'(L)^*\phi(L)\\
&-& \psi(r_H)^*\phi'(r_H) + \psi'(r_H)^*\phi(r_H)=0\,,
\label{symmetric1}
\end{eqnarray}
where $\psi(z), \phi(z) \in D(\tilde H_{BH})$.
We have assumed that the field goes to zero  asymptotically at a large
distance $\sim L$. It can be easily seen in the following way. For
large distance the effective radial differential equation, after
neglecting terms of $\mathcal{O}(1/z)$ and lower in Eq.(\ref{loperator}), becomes
\begin{eqnarray}
\frac{d^2\mathcal{F}_\omega(z)}{dz^2}= (m^2_S- \omega^2)\mathcal{F}_\omega(z)\,,
\end{eqnarray}
whose decaying solution  is of the form
\begin{eqnarray}
\mathcal{F}_\omega(z)= e^{-\sqrt{m_S^2-\omega^2}~z}\,.
\end{eqnarray}
Note that the quantity ($m_S^2-\omega^2$) under square root is
positive for bound state solutions, and $L^{-1}\simeq
\sqrt{(m_S^2-\omega^2)}$. Therefore the condition
(\ref{symmetric1}) reduces to
\begin{eqnarray}
\nonumber
&& \int_{r_H}^{L}\psi(z)^* \tilde H_{BH}\phi(z)dz - \int_{r_H}^{L}\left(\tilde H_{BH}\psi(z)\right)^* \phi(z)dz\\
&=&  \psi'(r_H)^*\phi(r_H) -\psi(r_H)^*\phi'(r_H)=0\,,
\label{symmetric2}
\end{eqnarray}
which depends only on short distance cutoff $r_H$. The  boundary condition consistent with (\ref{symmetric2}) will then  be
\begin{eqnarray}
D(\tilde{H}_{BH}) = \{\psi(z): \psi(r_H)= 0\}\,,
\label{boundary1}
\end{eqnarray}
which makes the operator self-adjoint too \cite{reed}.
As mentioned above, the field action corresponding to the operator
(\ref{latz0}) is  scale invariant under the transformation $t\to
\alpha^2 t$ and $z\to \alpha z$, where $\alpha$ is the scale
factor. The scale invariance implies that if $\chi(z)$  is an
eigen-state of the operator  $\tilde{H}_{BH}$ with eigenvalue
$\epsilon_{BH}$ then $\chi_\alpha= \chi(\frac{z}{\alpha})$ is also an
eigen-state of the same operator with eigenvalue
$\frac{\epsilon_{BH}}{\alpha^2}$. Note that the eigenvalue
$\epsilon_{BH}$ has nothing to do with the scalar field
eigenvalue $\omega$.  In general $\alpha$ is a continuous
parameter, which means the negative energy ground state will become
unstable for $\alpha \to 0$. One way to avoid this problem is to
ask for the possible self-adjoint extensions \cite{reed} which then
can make the
ground state stable. The self-adjoint extension parameter is related to the
boundary condition. In our problem of
black hole spacetime, however, black hole area quantization from
quantum gravity effects may eventually break the continuous scale
symmetry of the problem. As conjectured by Bekenstein \cite{beke, beke2} the
black hole horizon area  $A=4\pi r_+^2$ has a discrete spectrum
\begin{eqnarray}
\mathcal{A}_n= \epsilon \hbar n, n\in \mathbb{N}^+ \,,
\label{bek}
\end{eqnarray}
when it is treated as an  operator in quantum gravity theory. In Eq.(\ref{bek})
above, we have used units $G=c=1$, so that the Planck length
$l_P\sim\hbar^{1/2}$, and the Planck area is proportional to
$\hbar$. The quantity $\epsilon \hbar$ is the constant spacing between
the consecutive levels.
The radius of the horizon goes like
${r_+}_n \sim \sqrt{n}$ through  area quantization for $D=3$.
We exploit this quantization while making the scale transformation
\begin{eqnarray}
\alpha z= \tilde r - {r_+}_m= \alpha r-\alpha {r_+}_n\,.
\label{length1}
\end{eqnarray}
From the second and third equality we see multiplying $\alpha$
to ${r_+}_n$ will take to a different  ${r_+}_m$ only if $\alpha$ is
quantized, implying that scale symmetry is broken into discrete scale
symmetry by the area quantization. In general  (\ref{length1})
gives rise to the condition
\begin{eqnarray}
\alpha_m= \sqrt{m}\,, m\in \mathbb{N}^+
\label{lengthcond1}
\end{eqnarray}
The above scale transformation is rather ugly because it is different
for each spatial dimension $D$. It is more satisfactory, instead, to
take the scale transformation to be
\begin{eqnarray}
\alpha_m={m}\,, m\in \mathbb{N}^+
\label{lengthcond2}
\end{eqnarray}
which is valid for general space dimension $D$.

Let us now consider the exact differential equation for the scalar
field given by  (\ref{radial}).  Solutions of this equation which are
globally valid are not scale invariant. However  the near horizon
limits of the solutions are  certainly scale invariant. Suppose
$\mathcal{F}_{1\omega}(z)$ and $\mathcal{F}_{2\omega}(z)$ are two
independent solutions of the
differential equation (\ref{radial}). The general solution would be
of the form
\begin{eqnarray}
\mathcal{F}_{\omega}(z)= B_1\mathcal{F}_{1\omega}(z) +
B_2\mathcal{F}_{2\omega}(z)~.
\end{eqnarray}
Due to  short distance scale invariance we can construct a set
of wavefunctions of the form
$\mathcal{F}_{\omega}(\alpha_m^{-1}z), \alpha_m\in \mathbb{N}^+$. Although
$\mathcal{F}_{\omega}(\alpha_m^{-1}z)$ are not solutions of
(\ref{radial})
we know that the short distance limit
$\lim_{z\to
0}\mathcal{F}_{\omega}(\alpha_m^{-1}z)
\simeq\tilde{\mathcal{F}}_{\omega}(\alpha_m^{-1}z)$
are solutions  near the event horizon. Note that usually the scale
transformation $z\to \alpha z$ takes  the wave-function function
$\psi(z)\to \psi(z/\alpha)$.
We then  take these solutions and make a linear sum of them to impose a
symmetric  boundary condition of the form (\ref{boundary1}), which implies
\begin{eqnarray}
\sum_{\alpha_m=1}^{\infty} \tilde{\mathcal{F}}_{\omega}(\alpha_m^{-1}z)=0\,,
\label{bcond1}
\end{eqnarray}
at the brick wall. We have taken the same weight over the linear sum
since there is nothing to prefer one $m$ from another.
Therefore, the behavior of
$\tilde{\mathcal{F}}_{\omega}(\alpha_m^{-1}z)$
would be of the form
\begin{eqnarray}
\nonumber &&\tilde{\mathcal{F}}_{\omega}(\alpha_m^{-1}z)= \\
&&C_1(m^{-1}r_H)^{1/2 +iE}+ C_2(m^{-1}r_H)^{1/2 -iE}\,,
\end{eqnarray}
where $C_1, C_2$ are some constants which are not important for our
present purpose.  The boundary condition (\ref{bcond1}) then implies
\begin{eqnarray}
\zeta(1/2+iE)C_1 r_H^{1/2 +iE}
  + \zeta(1/2- iE)C_2 r_H^{1/2 -iE}=0
\label{bcond2}
\end{eqnarray}
The scalar field eigenvalues which are independent of short distance
cutoff can be obtained from (\ref{bcond2}) by separately making the
coefficients  zero, which implies that  eigenvalue equation for the
scalar field  is
\begin{eqnarray}
\zeta(1/2+i E)= 0\,.
\end{eqnarray}
The scalar field eigenvalues thus coincide with the nontrivial zeros
of the Riemann zeta function.

{\it V. Discussion.---}
Although we considered the background  to be a $(3+1)$-dimensional
Schwarzschild  spacetime, the analysis is applicable to a more
generic black hole spacetime, because the effective Hamiltonian
for the  field in the near-horizon region for a class of black holes are
known to have the   scale invariant form \cite{suneeta}.

Note that in \cite{berry}, a similar discrete scaling was made
and a sum taken over the
scaled states of the symmetrized operator $D=xp$. Then the zeta function
$\zeta(1/2+iE)$ came out as the overall coefficient of the resulting
state. Unlike in our case, at the zeros of the zeta function,
therefore, the wave function $\Psi(x)$ was a null state for all $x$. \\


{\it Acknowledgement.---}
P. R. Giri acknowledges the hospitality of the Department of Physics and Astronomy of  McMaster University. This work is supported by a grant of NSERC of Canada. Thanks are due to R. K. Bhaduri, J. Law, M. V. N. Murthy and D. W. L. Sprung for going through the first version of the manuscript carefully.

\end{document}